\documentclass[]{article}
\usepackage{lmodern}
\usepackage{amssymb,amsmath}
\usepackage{ifxetex,ifluatex}
\usepackage{fixltx2e} 
\ifnum 0\ifxetex 1\fi\ifluatex 1\fi=0 
  \usepackage[T1]{fontenc}
  \usepackage[utf8]{inputenc}
\else 
  \ifxetex
    \usepackage{mathspec}
  \else
    \usepackage{fontspec}
  \fi
  \defaultfontfeatures{Ligatures=TeX,Scale=MatchLowercase}
\fi
\IfFileExists{upquote.sty}{\usepackage{upquote}}{}
\IfFileExists{microtype.sty}{%
\usepackage[]{microtype}
\UseMicrotypeSet[protrusion]{basicmath} 
}{}
\PassOptionsToPackage{hyphens}{url} 
\usepackage[unicode=true]{hyperref}
\hypersetup{
            pdftitle={Data Readiness Levels},
            pdfauthor={Neil D. Lawrence},
            pdfborder={0 0 0},
            breaklinks=true}
\usepackage[margin=1in]{geometry}
\usepackage{color}
\usepackage{fancyvrb}

\DefineVerbatimEnvironment{Highlighting}{Verbatim}{commandchars=\\\{\}}
\newenvironment{Shaded}{}{}

\newcommand{\ImportTok}[1]{#1}

\newcommand{\OperatorTok}[1]{\textcolor[rgb]{0.40,0.40,0.40}{#1}}

\newcommand{\NormalTok}[1]{#1}
\IfFileExists{parskip.sty}{%
\usepackage{parskip}
}{
\setlength{\parindent}{0pt}
\setlength{\parskip}{6pt plus 2pt minus 1pt}
}
\ifx\paragraph\undefined\else
\let\oldparagraph\paragraph
\renewcommand{\paragraph}[1]{\oldparagraph{#1}\mbox{}}
\fi
\ifx\subparagraph\undefined\else
\let\oldsubparagraph\subparagraph
\renewcommand{\subparagraph}[1]{\oldsubparagraph{#1}\mbox{}}
\fi

\makeatletter
\def\fps@figure{htbp}
\makeatother

\title{Data Readiness Levels}
\author{Neil D. Lawrence\\Amazon Research Cambridge and University of Sheffield}
\providecommand{\institute}[1]{}
\institute{Amazon Research Cambridge and University of Sheffield}
\date{6th April 2017}

\begin{document}
\maketitle
\begin{abstract}
Application of models to data is fraught. Data-generating collaborators
often only have a very basic understanding of the complications of
collating, processing and curating data. Challenges include: poor data
collection practices, missing values, inconvenient storage mechanisms,
intellectual property, security and privacy. All these aspects obstruct
the sharing and interconnection of data, and the eventual interpretation
of data through machine learning or other approaches. In project
reporting, a major challenge is in encapsulating these problems and enabling
goals to be built around the processing of data. Project overruns can
occur due to failure to account for the amount of time required to
curate and collate. But to understand these failures we need to have a
common language for assessing the readiness of a particular data set.
This position paper proposes the use of data readiness levels: it gives
a rough outline of three stages of data preparedness and speculates on
how formalisation of these levels into a common language for data
readiness could facilitate project management.
\end{abstract}

\section{Introduction}\label{introduction}

The universal formula of machine learning is \[\text{data} +
\text{model} \rightarrow \text{prediction}\] with a high quality of
prediction being dependent on both good models and high quality data.
Much of the focus on machine learning in the academic world of machine
learning is on the quality of the model. This focus arises because data
is normally from benchmarks or publicly available data sets, so
performance in a task is improved by exercising control over model.

The greater interconnectedness of modern society and apparent ease with
which we digitise or record data places us in a particular position. We
need to develop a lot more control over the quality of our data. Whether
it be in the manner in which we choose to collect, or how we choose to
annotate. Currently the academic or empirical study of modeling
(e.g.~support vector machines, neural networks, Gaussian processes) is
prominent in the education of the graduates we produce, but approaches
for understanding the quality of data are less widely used.

\section{Data Readiness Levels}\label{data-readiness-levels}

In this position paper we introduce the idea of \emph{data readiness
levels}. Data readiness levels are designed to deal with a challenge for
human cognitive information processing. It's difficult for us to reason
about concepts when we haven't developed a language to describe them.
The idea of data readiness levels is to correct this issue and make it
easier for us to reason about the state of our data.

The challenges of data quality arise before modeling even starts. Both
questions and data are badly characterized. This is particularly true in
the era of Big Data, where one gains the impression that the depth of
data-discussion in many decision making forums is of the form ``We have
a Big Data problem, do you have a Big Data solution?'', ``Yes, I have a
Big Data solution.'' Of course in practice it also turns out to be a
solution that requires Big Money to pay for because no one bothered to
scope the nature of the problem, the data, or the solution.

Data scientists and statisticians are often treated like magicians who
are expected to wave a model across a disparate and carelessly collated
set of data and with a cry of `sortitouticus' a magical conclusion is
drawn. It is apt to think of it as ocean of data, to paraphrase ``water,
water everywhere and not a drop to drink'', we have ``data, data
everywhere and not a set to process''. Just as extracting drinkable
water from the real ocean requires the expensive process of
desalination, extracting usable data from the data-ocean requires a
significant amount of processing.

For any data analyst, when embarking on a project, a particular
challenge is assessing the quality of the available data, how much
processing is required? This difficulty can be compounded when project
partners do not themselves have a deep understanding of the process of
data analysis. If partners are not data-savvy they may not understand
just how much good practice needs to be placed in the curation of data
to ensure that conclusions are robust and representative. Just as water
quality should be measured before consumption, so should data quality.

When scoping a project, in most proposal documents, very scant attention
is paid to these obstacles, meaning in practice the process of improving
data quality is under appreciated and under resourced.

One difficulty is that the concept of ``data'', for many people, is
somehow abstract and diffuse. This seems to mean that it is
challenging for us to reason about. Psychologists refer to the idea of
\emph{vivid} information as information that is weighted more heavily in
reasoning than non-vivid or \emph{pallid information}. In this sense
data seems to be rendered \emph{vivid} to be properly accounted for in
planning. This may relate to availability heuristics{[}1{]}.

This abstract nature is also true of other terms, for example for many
people the idea of ``technology'' is also similarly diffuse, it is
pallid information (although I've never heard anyone remark ``we have a
big technology problem, we need a big technology solution''). Perhaps to
deal with this challenge, in large scale projects, when deploying
technology, we are nowadays guided to consider its \emph{readiness
stage}. Technology readiness levels arose in NASA {[}2{]}. The readiness
of the technology is made manifest through a set of numbers which describe its
characteristics:\footnote{See appendix for examples of technology
  readiness level descriptions.} is it lab tested only? Is it ready for
commercialization? Is it merely conceptual?

No doubt there are pros and cons of such readiness levels, but one of
the pros is that the manifestation of the technological readiness
pipeline ensures that some thought is given to that process. The
technology is rendered more vivid even through a shared
characterization.

It would therefore seem very useful to have a scale to make \emph{data
readiness} manifest. This idea would allow analysts to encourage better
consideration of the data collection/production and consolidation, with
a set of simple questions, ``And what will the data readiness level be
at that point?''. Or ``How will that have progressed the data
readiness?''. Or to make statements, ``we'll be unable to deliver on
that integration unless the data readiness level is at least
B3..''\footnote{The nanotechnology community has also looked at data
  readiness levels in this
  \href{http://www.nano.gov/node/1015}{discussion document from the
  nanotechnology community in 2013}. However, their scope doesn't seem
  to be general enough to deal with the challenges of data processing in
  domains beyond nanotechnology.}

This paper aims to trigger a discussion in statistics and data science
communities by proposing an initial set of descriptors for data
readiness.

The initial proposal is that data readiness should be split into three
\emph{bands}. Each band being represented by a letter, A, B and C. These
bands reflect stages of data readiness which would each likely have some
sub-levels, so the best data would be A1 and the worst data might be C4.
The aim here is to avoid being fine-grained too early. We therefore
begin the discussion by focussing on three bands of data readiness.

\subsection{Band C}\label{band-c}

Band C is about the accessibility of a data set. The lowest sub-level of
Band C (let's label it as C4) would represent a belief that the data may
exist, but its existence isn't even verified. Signs that data is C4
might include statements like ``The sales department should have a
record of that.'' Or ``The data should be available because we
stipulated it in the software requirements.'' We might think of it as
\emph{hearsay} data. Data that you've heard about so you say it's there.
Problems with hearsay data might include

\begin{itemize}
\item
  whether it really is being recorded
\item
  the format in which it's being recorded (e.g.~handwritten log book,
  stored in PDF format or old machine formats)
\item
  privacy or legal constraints on the accessibility of the recorded
  data, have ethical constraints been alleviated?
\item
  limitations on access due to topology (e.g.~the data's distributed
  across a number of devices)
\end{itemize}

So when we are first told a data set exists, when we have hearsay data,
then it is at band C4. For data to arrive at C1, then it would have all
these considerations dealt with.

When data arrives at C1 it's ready to be loaded into analysis software,
or it can be made available for others to access (e.g.~via a data
repository such as OpenML {[}3{]}). It is machine readable and ethical
procedures for data handling have been addressed. Bringing data to C1 is
often a significant effort itself involving many lines of bespoke
software and human understanding of systems, ethics and the law.

Some parts of Band C are sometimes referred to as ``data munging'' or
``data wrangling'', but those aren't the only components of this band,
there are additional challenges such as ethical and legal that need to
be resolved.

\subsection{Band B}\label{band-b}

Band B is about the faithfulness and representation of the data. Now
that it's loaded into the software, is what is recorded matching what is
purported to be recorded? How are missing values handled, what is their
encoding? What is the noise characterization (for sensors) or for manual
data are there data entry errors? Are any scientific units correctly
formulated?

Tukey's approach of ``Exploratory Data Analysis'' also fits within Band
B. Visualizations of the data should be carried out to help render the
data vivid and to ensure decision makers, who may not be data aware, can
become involved in the analysis process. Decision makers (e.g.~project
managers, or the client) should also begin to get a sense of the
limitations of their data set through appropriate visualisation.

As part of Band B the characteristics of the collection process should
also be verified, was data collection randomized, is it biased in any
particular way?

Other things to watch for at this stage include:

\begin{enumerate}
\def\labelenumi{\arabic{enumi}.}
\item
  If the data has been agglomerated at some point (for example, for
  privacy) how were missing values dealt with before agglomeration? If
  they weren't dealt with then that entire section of the data may be
  invalidated
\item
  If the data has been through a spreadsheet software, can you confirm
  that no common spreadsheet analysis errors were made? For example, was
  a column or columns accidentally perturbed (e.g.~through a sort
  operation that missed one or more columns)? Or was a
  \href{http://bmcbioinformatics.biomedcentral.com/articles/10.1186/1471-2105-5-80}{gene
  name accidentally converted to a date}?
\end{enumerate}

By the end of Band B, when data is B1, a broad idea of limitations in
the data should be present in the expert's mind. Data at C4 was hearsay
data, someone heard the data existed and they said what they thought it
might be good for. At B1 the analyst knows how faithful the data is to
that description. This is the significant challenge for a data
scientist. What people believe they have in their data versus what's
actually there. Only at the end of Stage B would the analyst begin to
have an intuition about what my really be possible with the data set.
Getting to this point is often the most expensive part of the project,
but we do not yet have good methods to guage progress, or share the
status of a particular data set.

\subsection{Band A}\label{band-a}

Band A is about data in context. It is at Band A that we consider the
appropriateness of a given data set to answer a particular question or
to be subject to a particular analysis.

The context must be defined. For example OpenML {[}3{]} defines
\emph{tasks} associated with data sets. A data set can only be
considered in Band A once a task is defined. A task could be ``Use the
data to predict a user preference'' or ``Use this data to prove the
efficacy of a drug'' or ``Use this data verify the functioning of our
rocket engine''.

Once data has been considered alongside a task and any remedial steps
have been taken, then the data is in A1 condition. It is ready to be
deployed in the context given and it can be used to make predictions
with the data.

Because A1 is about data in context, it is possible for a data set to be
A1 for one question (e.g.~predicting customer churn) but only B1 for a
different question (e.g.~predicting customer susceptability to a
particular special offer). So the definition of the context or task is
an important pre-requisite for this band.

To bring a data set up to A1, there may be a need for significant
annotation of the data by human expert. There may be a realisation that
new data needs to be actively collected to get the answers required. In
that sense, Band A has some characteristics of a classical statistical
analysis where the question would normally preceed the data collection.
It is in Band A where you should be carefully thinking about the
statistical design because it is only when you have the question you
wish to answer that you can really unpick how your data may be biased or
what information is missing.

\section{The Analysis Pipeline}\label{the-analysis-pipeline}

A common mistake in data analysis is to not acknowledge the different
processes above. They have different pre-requisites and require
different skill sets to carry out. However, this path cannot be
completely disconnected. Anyone performing an analysis at Band A also
needs to be intimately familiar with the collection process so that any
biases in data collection can be understood. Sharing information about
decisions taken at Bands B and C will also be critical to achieving a
good result.

What happens if we bring two data sets together to form a new data set?
Some assesment of data readiness would still need to be performed on the
new data set, even if the two other data sets had already been assessed.
Why? Well, one example is that there may be ethical issues with
combining data sets. For example, medical data can be easily
deanonymized if it is combined with other related data sets. That would
mean that some assesment would need to be made at Band C to see if it is
ethically responsible to combine the two data sets in the same analysis
package. Futher assesments would also need to be done at Band B to
understand differences in collection protocols that might make the two
data sets hard to combine.

\subsection{Potential Results}\label{potential-results}

The idea of these levels is to increase the accountability of the
process and allow the nature of the data to be manifest. With
data readiness levels in place you can now imagine conversations that
would include statements like the following:

\begin{quote}
Be careful, that department claims to have made 10,000 data sets
available, but we estimate that only 25\% of those data sets are
available at C1 readiness.
\end{quote}

\begin{quote}
The cost of bringing the data to C1 would be prohibitive for this study
alone, but the company-wide data audit is targeting this data to be C1
by Q3 2017 which means we can go ahead and recruit the statisticians we
need.
\end{quote}

\begin{quote}
The project failed because we over recruited statistical expertise and
then deployed them on bringing the data set to C1 readiness, a job that
would have been better done by building up our software engineering
resource.
\end{quote}

\begin{quote}
What's the data readiness level? My team will be ineffectual until it's
B1 and at the moment I see no provision in the plan for resource to
bring it there.
\end{quote}

\begin{quote}
We estimate that it's a \$100,000 dollar cost to bring that data to B1,
but we can amortize that cost across four further studies that also need
this data.
\end{quote}

\begin{quote}
I gave them the data readiness levels to go through and they realized
they hadn't yet got the necessary ethics approval for sharing the zip
codes. We'll revisit when they've got through that and can assure us
they can share a C1 set.
\end{quote}

\begin{quote}
While their knowledge of the latest methodologies wasn't as good as I'd
hoped, the candidate had a lot of experience of bringing data from C1 to
B1, and that's a skill set that we're in dire need of.
\end{quote}

\begin{quote}
The project came in under budget because they found a team with
experience of getting a closely related data set to A1. Many of the
associated challenges were the same and they could even reuse some of
that team's statistical models.
\end{quote}

\section{Case Studies}\label{case-study}

How useful would data readiness levels be? That's difficult to give a
quantitative answer to, large scale analysis of the scale of this
problem requires meta data science, i.e.~the study of data science
itself through acquiring data about data science. That would be a very
worthwhile endeavour. In the absence of quantitative information, we
provide two anecdotal case studies below. The first is on the cahllenge
of extracting data from a popular machine learning conference
proceedings.

\subsection{Proceedings of Machine Learning
Research}\label{proceedings-of-machine-learning-research}

The Proceedings of Machine Learning Research\footnote{http://proceedings.mlr.press}
(PMLR) were begun in 2006 (as JMLR Workshop and Conference Proceedings)
to provide a convenient way to publish machine learning conference
proceedings without the overhead of a conventional publisher.

There are now 69 volumes of proceedings published and planned. They
contain over 3,198 papers. The original website was manually curated to
mimic the JMLR website, but since Volume 26 an automated proceedings
production process which relies on editors providing a zip file of PDFs
and a bibtex bibliography reference file (referred to below as `bib
files') specifying author names, abstracts and titles has been used.

In early 2017, as part of a rebranding process, the original website was
moved to github and a new process for receiving proceedings and
publishing was set up. For the rebranding the old web-sites needed to be
scraped, converted to bib files.

A key aim in the rebanding was to make abstracts and titles easily
available for analysis with an initial target language of Python.

This entire process can be seen as taking the data readiness of website
data from C4 to C1. In other words, get to the point where the data
could be loaded into an analysis software, our target here was to load
it into the pandas framework within Python.

The \emph{original} plan was to complete the work in some idle hours at
the weekend. The actual work took much longer than projected. The major
github commits of code along with a description of the effort involved
are listed below.

\begin{enumerate}
\def\labelenumi{\arabic{enumi}.}
\item
  An initial two day effort to create bibliography files for the first
  26 volumes which were published before the publication process was
  first automated. Github commit:
  \url{https://github.com/mlresearch/papersite/commit/daa51a0da82b5e532214a732a3d32d94ddf66381}
\item
  A three day effort to convert the proceedings to a new format website.
  Much of this work was on the website presentation, but part was on
  data curation. Github commit:
  \url{https://github.com/mlresearch/papersite/commit/81d7a0556948281d11d8f0652ecdca61005c4318}
\item
  A three evening effort to tidy the resulting data so it could load
  into Python pandas via web download. Github commit:
  \url{https://github.com/mlresearch/papersite/commit/81d7a0556948281d11d8f0652ecdca61005c4318}
\end{enumerate}

The result is now available via github as a short jupyter notebook:
\url{https://github.com/sods/ods/blob/master/notebooks/pods/datasets/pmlr_data.ipynb}.
Data loading is now possible with a single library call.

\begin{Shaded}
\begin{Highlighting}[]
  \ImportTok{import}\NormalTok{ pods}
\NormalTok{  data }\OperatorTok{=}\NormalTok{ pods.datasets.pmlr()}
\end{Highlighting}
\end{Shaded}

where the pods library is available from
\url{https://github.com/sods/ods/}.

This data has therefore successfully transferred from the start of C
(e.g.~C4, ``Hearsay data'') to the end of C (e.g.~C1, data loaded into
analysis software).

The data set only contains 3,198 data points, it was all available in
electronic form. There were no issues around privacy or intellectual
property but work still took approximately six working days. Much of
that work was laborious, but it still involved well qualified
understanding of computer software, e.g.~regular expressions, scripting
languages, libraries for downloading from the internet, github etc..
Alongside that work a new format web page was also provided
(\url{http://proceedings.mlr.press}). Indeed that could be argued as a
consequence of the data tidy up. Naturally the hope is that future
volumes will be more cleanly added to this data set.

\subsection{Data Wrangling Snafus}\label{data-wrangling-snafus}

Some snafus that occurred in the data wrangling of the Proceedings of
Machine Learning Research.

Each bibliography file is provided by the editors of the relevant
volume. Because they each used different approaches to generate the bib
files, there were different issues with each bib file.

\begin{itemize}
\item
  Many of these bib files will have been produced from conference
  management software (e.g.~CMT or Easychair). That means the original
  information source is often likely to be author provided. Many authors
  seemed to paste their abstracts from the PDFs of their papers into
  this software. This meant that abstracts contained ligatures. A
  ligature is a single typeset unit such as `fi', which comes about when
  the previous letter runs into the second, the German `Sz', ß, is an
  example of a ligature that in the end became a letter. Unfortunately
  the ligatures did not paste as unicode, but as escaped characters
  which the python yaml library was unable to read when presented as
  data.
\item
  Papers are mostly written in Latex, so many of the abstracts contain
  Latex commands. But these commands in the abstracts or title are
  intepreted as escaped characters in yaml files. Additional `escaping'
  was needed for these commands.
\item
  Author names containing accents were oftentimes corrupted, perhaps due
  to differences in representation in the original files.
\end{itemize}

Each of the snafus above can be resolved by robust coding, but there is
normally a trade off between producing the data (just getting it done
quick) and producing the most robust code (ensuring the result is high
quality and reusable). The right operating point for the trade off is
driven by the scale of the data set. It also requires experience to
judge and is dependent on the coding skills of the data engineer.

This is the work at the pit face of data mining, it's difficult to
estimate the time it will take, and yet it is normally a critical
dependency in any machine learning project. For each of these snafus,
better planning at early stage analysis could have saved time in manual
correction of the data later. But doing such planning well requires
experience and an understanding of best practice. We are not associating
enough value with this experience, and therefore we continue to be
tripped by snafus when preparing data.

\subsubsection{How Data Readiness Levels could have
helped}\label{how-data-readiness-levels-could-have-helped}

This is one of the first projects I undertook after conceiving of data
readiness levels, but in reflection I think I still did not take enough
of the ideas seriously enough. Because I underestimated the time to be
spent on manual curation after the initial scrape of the data, I did not
put enough effort into robust code for dealing with ligatures and other
unusual character codes.

There were several moments where, in retrospect, I should have
refactored my processing code. But since I was driven by the desire to
complete the project, particularly given I'd severely underestimated how
long it would take, I chose to push forward. One interesting question is
how ideas from software engineering, such as agile development
philosophies, could have helped in making myself more aware of these
errors.

\subsection{Disease Monitoring in
Uganda}\label{disease-monitoring-in-uganda}

The second case study refers to work in collaboration with the
University of Makerere and UN Global Pulse in Kampala, Uganda in the
prediction of disease outbreak {[}4{]}. Specifically, our original
interest was understanding the spatial correlations of malarial disease
in Uganda, and their interactions with other measures such as NDVI,
rainfall, altitude etc. Our effort was a countrywide effort, but
simultaneous to our work was an international effort in the Malaria
Atlas Project.\footnote{http://www.map.ox.ac.uk/}

We constructed spatial models based on Gaussian processes that were
designed to detect and use the correlations between the different
covariates. The work was done through two PhD students' thesis, Martin
from Kampala, and Ricardo from the UK. The work in Kampala mainly drove
the data collection and collation. In particular, interpretation of
satellite images as NDVI, alignment of the rainfall and altitude maps.
The data collation in itself took probably around 70-80\% of the work on
the project. Martin working on it full time, and Ricardo assisting and
working on both modeling and data munging. Ricardo visited Uganda twice,
initially for a scoping visit and later for a longer collaboration visit
in the processing of the data during his PhD. These visits were to the
University of Makerere where Martin was working.

The malaria incidence data was drawn from Health Management Information
System (HMIS) data. For privacy purposes, original data was not
available to us, but rather it had been aggregated across certain
regions.

One key challenge in the data munching is that the administrative areas
(in Uganda known as districts) change. One administrative area can
divide into two. When this happens the history of the two districts
needs to be divided across the two areas. This presented a missing data
problem that Ricardo also spent a deal of time addressing during his
thesis.

By the time of submission of the thesis, Ricardo had recovered a
negative result, he couldn't show spatial correlation among the
districts. This was disappointing as it had been one of the main
motivations of that thesis work. Nevertheless, even without the spatial
correlation there was interest in deploying the system by the Ugandan
ministry of health for disease outbreak prediction.

Ricardo's expertise was such that he was able to then spend 3 months in
Africa, this time at UN Global Pulse, where Martin and his supervisor
were now working, rather than Makerere. This was to help with the
implementation of early warning systems based on the model. These early
warning systems did not require the spatial correlation that hadn't
worked during the thesis.

During the period at the UN, Ricardo and Martin were able to work
directly with disaggregated data. The UN had permission to see the raw
health center results, whereas the University of Makerere did not. As a
result, they found that aggregation we had previously been working on
had been performed across data containing missing values. Ricardo and
Martin reworked the model to deal with missing values and recovered the
expected spatial correlation.

In other words, due to a data processing error at Band B, a negative
result was obtained at Band A. A lack of documentation, or a lack of
asking the right questions, led to an oversight on this error.

The models are now deployed for disease outbreak prediction across
Uganda, where they are combined with knowledge of population movement
from mobile phone data to trigger interventions.

The lessons learned were the following.

\begin{enumerate}
\def\labelenumi{\arabic{enumi}.}
\item
  In data where we had direct access the amount of work required to
  align, and the number of design decisions we made in the summarization
  dominated the project.
\item
  There was still data that fell out of our control due to
  confidentiality reasons. Despite the work we'd done on our own data,
  which would have allowed us to infer that other design decisions would
  have been made by those that were aggregating the data, we failed to
  ask the right questions of the data providers. Indeed, those data
  providers may not have even known the answers. There was no substitute
  for directly looking at the data.
\item
  Although we began the project with one set of goals in mind:
  understanding the correlation between different factors in malaria,
  our outcome: disease prediction for early intervention, was quite
  different. Happenstance outcomes need to be accommodated in project
  goals. They arise particularly around Band B of the process.
\end{enumerate}

\subsubsection{How Data Readiness Levels could have
helped}\label{how-data-readiness-levels-could-have-helped-1}

Being aware of data readiness levels would have done three things.
Firstly, we would have estimated better how much time that Martin would
require in data munging from satellite images etc. Secondly, we would
have questioned the process by which the HMIS data was being presented
to us, and what stage of readiness it was at, and how it got there.
These are broadly questions that sit in Band C and B. Thirdly, awareness
of the transition from Band B to Band A (data moving to its context)
would have made us realize that the question may well evolve and be more
responsive to that outcome. In the end this move was driven by the shift
in collaboration from Makerere to the UN.

Ricardo has continued to work in this area, and one focus of his new
UCSF Global Health research group has been to clean up existing data and
make it available as geospatial layers for other groups to use.

\section{Conclusion}\label{debate}

Machine learning researchers probably didn't enter the field to do
project management, but it may be that many failings on large data
projects are associated with a failure to provision resource for the
challenges involved in preparing our data, rather than a failing in the
algorithmics of the system. Costs of data curation are often
underestimated and those who do the work in Band C and Band B are very
often undervalued.

Data readiness levels highlight the different skill sets required in
each stage of analysis, from software engineer, to data-munger, to data
scientist to machine learning scientist.

Some consensus about such levels would help organizations (and their
managers, financial controllers) quantify the value associated with data
and allocate resource correctly to developing data sets that are robust
and representative. A well conducted data analysis will lead to a good
customer experience, but by the same token badly waste resources and
give a poor customer experience.

This paper has outlined the basis for a partial solution to these
problem, but to deploy in practice we would need consensus around data
readiness levels and how they should be deployed. The emerging field of
data science is the ideal domain to explore the utility of these ideas,
evolve their specification and begin to properly account for the value
of well curated data.

For convenient reference below we show examples of \emph{technology}
readiness levels from the Department of Defense (sourced from
\href{https://en.wikipedia.org/wiki/Technology_readiness_level}{Wikipedia}.
For \emph{Data} Readiness we are proposing to start with the three bands
described above. Technology readiness relies entirely on a numbered
system.

\hypertarget{refs}{}
\hypertarget{ref-Gilovich:heuristics02}{}
{[}1{]} T. Gilovich, D. Griffin, and D. Kahneman, Eds., \emph{Heuristics
and biases: The psychology of intuitive judgment}. Cambridge University
Press, 2002.

\hypertarget{ref-Banke:trl10}{}
{[}2{]} J. Banke, ``Technology readiness levels demystified,'' NASA,
2010.

\hypertarget{ref-pmlr-v41-vanschoren15}{}
{[}3{]} J. Vanschoren, J. N. Rijn, and B. Bischl, ``Taking machine
learning research online with OpenML,'' in \emph{Proceedings of the 4th
international workshop on big data, streams and heterogeneous source
mining: Algorithms, systems, programming models and applications}, 2015,
vol. 41, pp. 1--4.

\hypertarget{ref-Andrade-Pacheco2016}{}
{[}4{]} R. Andrade-Pacheco, M. Mubangizi, J. Quinn, and N. Lawrence,
``Monitoring short term changes of infectious diseases in uganda with
gaussian processes,'' in \emph{Advanced analysis and learning on
temporal data: First ecml pkdd workshop, aaltd 2015, porto, portugal,
september 11, 2015, revised selected papers}, A. Douzal-Chouakria, J. A.
Vilar, and P.-F. Marteau, Eds. Cham: Springer International Publishing,
2016, pp. 95--110.

\appendix

\section{Exemplar Technology Readiness
Levels}\label{exemplar-technology-readiness-levels}

\begin{enumerate}
\def\labelenumi{\arabic{enumi}.}
\item
  Basic principles observed and reported

  Lowest level of technology readiness. Scientific research begins to be
  translated into applied research and development (R\&D). Examples
  might include paper studies of a technology's basic properties.
\item
  Technology concept and/or application formulated

  Invention begins. Once basic principles are observed, practical
  applications can be invented. Applications are speculative, and there
  may be no proof or detailed analysis to support the assumptions.
  Examples are limited to analytic studies.
\item
  Analytical and experimental critical function and/or characteristic
  proof of concept

  Active R\&D is initiated. This includes analytical studies and
  laboratory studies to physically validate the analytical predictions
  of separate elements of the technology. Examples include components
  that are not yet integrated or representative.
\item
  Component and/or breadboard validation in laboratory environment

  Basic technological components are integrated to establish that they
  will work together. This is relatively ``low fidelity'' compared with
  the eventual system. Examples include integration of ``ad hoc''
  hardware in the laboratory.
\item
  Component and/or breadboard validation in relevant environment

  Fidelity of breadboard technology increases significantly. The basic
  technological components are integrated with reasonably realistic
  supporting elements so they can be tested in a simulated environment.
  Examples include ``high-fidelity'' laboratory integration of
  components.
\item
  System/subsystem model or prototype demonstration in a relevant
  environment

  Representative model or prototype system, which is well beyond that of
  TRL 5, is tested in a relevant environment. Represents a major step up
  in a technology's demonstrated readiness. Examples include testing a
  prototype in a high-fidelity laboratory environment or in a simulated
  operational environment.
\item
  System prototype demonstration in an operational environment.

  Prototype near or at planned operational system. Represents a major
  step up from TRL 6 by requiring demonstration of an actual system
  prototype in an operational environment (e.g., in an aircraft, in a
  vehicle, or in space).
\item
  Actual system completed and qualified through test and demonstration.

  Technology has been proven to work in its final form and under
  expected conditions. In almost all cases, this TRL represents the end
  of true system development. Examples include developmental test and
  evaluation (DT\&E) of the system in its intended weapon system to
  determine if it meets design specifications.
\item
  Actual system proven through successful mission operations.

  Actual application of the technology in its final form and under
  mission conditions, such as those encountered in operational test and
  evaluation (OT\&E). Examples include using the system under
  operational mission conditions.
\end{enumerate}

\end{document}